\documentclass[epj]{svjour}
\usepackage{graphics}
\newcommand{\be}{\begin{equation}}
\newcommand{\ee}{\end{equation}}
\begin{document}
\title{The shell model Monte Carlo approach to level densities: recent developments and perspectives}
\author{Y. Alhassid
}                     
%
%
\institute{Center for Theoretical Physics, Sloane Physics 
Laboratory, Yale University, New Haven, Connecticut 06520, USA}
\date{Received: 7 June 2015/ Revised: 2 December 2015}
%
\abstract{
We review recent advances in the shell model Monte Carlo approach for the microscopic calculation of statistical and collective properties of nuclei. We discuss applications to the calculation of (i) level densities in nickel isotopes, implementing a recent method to circumvent the odd-particle sign problem; (ii) state densities in heavy nuclei; (iii) spin distributions of nuclear levels; and (iv) finite-temperature quadrupole distributions. 
} 
\maketitle
\section{Introduction}
\label{intro}

The nuclear level density is among the most important statistical properties of nuclei and is key input in the Hauser-Feshbach theory~\cite{Hauser1952} of statistical nuclear reactions. However, its microscopic calculation in the presence of correlations is a challenging many-body problem. Most approaches are based on empirical models such as the back-shifted Fermi gas model and the constant temperature formula, and on mean-field theories such as the Hartree-Fock-Bogoliubov (HFB) approximation~\cite{Konig2008,Hilaire2012}. 

The configuration-interaction (CI) shell model approach offers an attractive framework that includes correlations beyond mean-field theory and shell effects, but its applications in mid-mass and heavy nuclei have been hindered by the combinatorial increase of the dimensionality of the many-particle model space with number of orbitals and  number of valence nucleons.  The shell model Monte Carlo (SMMC) method~\cite{Lang1993,Alhassid1994,Koonin1997,Alhassid2001}  enables calculations in model spaces that are many orders of magnitude larger than those that can be treated in conventional diagonalization methods, and is a powerful state-of-the-art method for the microscopic calculation of nuclear level densities~\cite{Nakada1997,Alhassid1999,Alhassid2007,Alhassid2008,Ozen2013}.  

Here we review recent advances in the SMMC method for calculating level densities. These include (i) a method to circumvent the odd-particle Monte Carlo sign problem~\cite{Mukherjee2012}, enabling the first accurate SMMC calculation of level densities in odd-mass nuclei~\cite{Bonett2013}; (ii) calculation of state densities in even-mass~\cite{Ozen2013} and odd-mass~\cite{Ozen2015} heavy nuclei; (iii) calculation of the spin distributions of level densities~\cite{Alhassid2007}, and (iv) a method to calculate the probability distribution associated with the quadrupole operator~\cite{Alhassid2014}, which might facilitate the modeling of level densities as a function of deformation, an important input for the calculation of fission rates. 

\section{Shell model Monte Carlo method}
\label{SMMC}

The SMMC method is based on the Hubbard-Stratonovich transformation~\cite{Hubbard1959,Stratonovich1957}, in which the Gibbs ensemble $e^{-\beta H}$ of a system described by a Hamiltonian $H$ at inverse temperature $\beta=1/T$ is expressed as a functional integral over one-body ensembles that describe non-interacting nucleons moving in external auxiliary fields  
\begin{equation}\label{HS}
e^{-\beta H} = \int {\cal D}[\sigma]
G_\sigma U_\sigma \;.
\end{equation}
 Here $G_\sigma$  is a Gaussian weight and $U_\sigma$ is the propagator of non-interacting nucleons moving in time-dependent auxiliary fields $\sigma(\tau)$ ($0\leq \tau \leq \beta$).   
 The thermal expectation value of an observable $O$ is calculated from
  \begin{equation}\label{observable}
 \langle O \rangle ={{\rm Tr}(e^{-\beta H}O) \over {\rm Tr} e^{-\beta H}} = {\int D[\sigma] W_\sigma \Phi_\sigma \langle O \rangle_\sigma
\over \int D[\sigma] W_\sigma \Phi_\sigma} \;,
 \end{equation}
where $W_\sigma = G_\sigma |{\rm Tr}\, U_\sigma|$ is a positive-definite weight function, $\Phi_\sigma = {\rm Tr}\, U_\sigma/|{\rm Tr}\, U_\sigma|$ is the Monte Carlo sign function, and $\langle O \rangle_\sigma =
 {\rm Tr} \,( O U_\sigma)/ {\rm Tr}\,U_\sigma$ is the thermal expectation value of the observable for a given configuration $\sigma$ of the auxiliary fields. 
  
The many-particle space is the Fock space spanned by a set of $N_s$ spherical single-particle orbitals. The many-particle propagator $U_\sigma$ can then be represented in the single-particle space by a matrix ${\bf U}_\sigma$ of dimension $N_s\times N_s$, and the quantities in the integrands of Eq.~(\ref{observable}) can be expressed in terms of this matrix. The grand canonical trace  of $U_\sigma$ is given by
\begin{equation}
{\rm Tr}\; U_\sigma = \det ( {\bf 1} + {\bf U}_\sigma) \;,
\end{equation}
while the grand canonical expectation value of a one-body observable $O =\sum_{i,j}  O_{ij} a^\dagger_i a_j$ can be calculated from
\be\label{1-body}
\langle a_i^\dagger a_j \rangle_\sigma
 = \left[ {1 \over {\bf 1} +{\bf U}^{-1}_\sigma}
\right]_{ji} \;.
\ee
The grand canonical expectation value of a two-body observable can be similarly calculated using Wick's theorem.
 
 In the finite nucleus it is important to calculate the observables at fixed numbers of protons and neutrons using the canonical ensemble. This is done using an exact representation of the particle-number projector as a Fourier sum. For example, the canonical partition function at fixed number of particles $A$ is given by~\cite{Ormand1994}
\begin{eqnarray}\label{canonical}
{\rm Tr}_A U_\sigma =\frac{e^{-\beta\mu  A}}{N_s}
\sum_{m=1}^{N_s} e^{-i\varphi_m A}\,\det \left({\bf 1}+e^{i\varphi_m}e^{\beta\mu}{\bf U}_\sigma\right)
\;,
\end{eqnarray}
where  $\varphi_m=2\pi m/N_s$ $(m=1,\ldots,N_s)$ are quadrature points and $\mu$ is a real chemical potential [required for the numerical stabilization of the sum in (\ref{canonical})].
 Similar expressions can be written for canonical expectation values of observables, relating them to grand canonical values through a Fourier transform.
 
 In SMMC, we choose configurations $\sigma_k$ of the auxiliary fields that are distributed according to $W_\sigma$ and then estimate the observables from
\begin{equation}
\langle O\rangle \approx  { {\sum_k  \langle  O \rangle_{\sigma_k} \Phi_{\sigma_k} \over \sum_k \Phi_{\sigma_k}}} \;.
\end{equation}

\subsection{State density in SMMC}
\label{density}
In SMMC, we calculate the thermal energy versus inverse temperature $\beta$ as the expectation value of the Hamiltonian $E(\beta)=\langle H\rangle$, and then integrate the thermodynamic relation  
\be
-{\partial \ln Z \over \partial \beta } = E(\beta)
\ee
to find the canonical partition function $Z(\beta)$.
The state density is determined from the partition function by an inverse Laplace transform
\be\label{inv-L} 
\rho(E) = {1 \over 2\pi i} \int_{-i \infty}^{i \infty} d\beta \,e^{\beta E} Z(\beta) \;. 
\ee 
 In practice we calculate the {\it average} state density by evaluating the 
integral in (\ref{inv-L}) in the saddle-point approximation~\cite{BM69}
\be\label{saddle} 
\rho(E) \approx \left(-2 \pi {d E \over d \beta}\right)^{-1/2}  
e^{S(E)}\;, 
\ee 
where $\beta$ is determined as a function of energy $E$ from the saddle-point condition $E(\beta)=E$, and  $S$ is the canonical entropy
\be
S = \ln Z +\beta E\;.
\ee 

\subsection{Circumventing the odd-particle sign problem}\label{Odd-A}

For good-sign interactions, the sign $\Phi_\sigma$ (in the grand canonical ensemble) is positive. The projection on an even number of particles keeps the sign of the projected partition function ${\rm Tr}_A U_\sigma$ positive. However, the projection on an odd number of particles can lead to negative sign of ${\rm Tr}_A U_\sigma$ for some of the samples. At low temperatures, this leads to the so-called sign problem where the statistical errors of the observables become too large. In particular, it is difficult to determine accurate ground-state energies of odd-mass nuclei. This odd-particle sign problem has hampered applications of SMMC to odd-mass nuclei. 

Recently, we developed a method to circumvent the odd-particle sign problem (for good-sign interactions) and determine accurate ground-state energies of odd-mass nuclei~\cite{Mukherjee2012}. The method is based on the asymptotic behavior of the single-particle Green's functions in imaginary time.  The scalar Green's functions are given by
\be\label{green}
G_{\nu}(\tau) = \frac{{\rm Tr}_{\mathcal{A}}\left[~e^{-\beta H} \mathcal{T} \sum_m a_{\nu m}(\tau) a^\dagger_{\nu m}(0)\right]}{{ \rm Tr}_{\mathcal{A}}~e^{-\beta H}}\;,
\ee
where $\nu \equiv (n l j)$ labels the nucleon single-particle orbital with radial quantum number $n$, orbital angular momentum $l$ and total spin $j$. Here $\mathcal{T}$ denotes
time ordering and $a_{\nu m}(\tau)\equiv e^{\tau H} a_{\nu m} e^{-\tau H}$ is an annihilation operator of a nucleon at imaginary time $\tau$ ($-\beta \leq \tau\leq \beta$) in a
single-particle state with orbital $\nu$ and magnetic quantum number $m$.

For an even-even nucleus $\mathcal{A}\equiv (Z,N)$ with ground-state spin 0, the scalar Green's function
has the asymptotic form  
\be
G_{\nu}(\tau) \sim e^{- \Delta E_{J=j}(\mathcal{A}_{\pm}) |\tau|} \;,
\ee
where $\mathcal{A}_\pm$ denote the even-odd nuclei $(Z,N \pm 1)$ when $\nu$ is a neutron orbital and the odd-even nuclei $(Z \pm 1,N)$ when $\nu$ is a proton orbital, and the $+$ ($-$) subscript should be used for $\tau > 0$ ($\tau \leq 0$). $\Delta E_{J=j}(\mathcal{A}_{\pm})$ is the difference between the energies of the the lowest spin $J$ eigenstate of the $\mathcal{A}_{\pm}$ particle nucleus and the ground state of the $\mathcal{A}$ particle nucleus.  In this asymptotic regime, we can thus calculate $\Delta E_j(\mathcal{A}_{\pm})$ from the slope of $\ln G_{\nu}(\tau)$. By minimizing $\Delta E_j(\mathcal{A}_{\pm})$ over all possible values of $J$, we determine the difference between the ground-state energy of the $\mathcal{A}_{\pm}$ nuclei, 
and the ground state energy of the $\mathcal{A}$ nucleus. This latter energy and $G_{\nu}(\tau)$ are both properties of the even-even nucleus and are free of a sign problem for a good-sign interaction.

\section{Application to nickel isotopes}

\begin{figure*}
\center\resizebox{0.85\textwidth}{!}{%
  \includegraphics{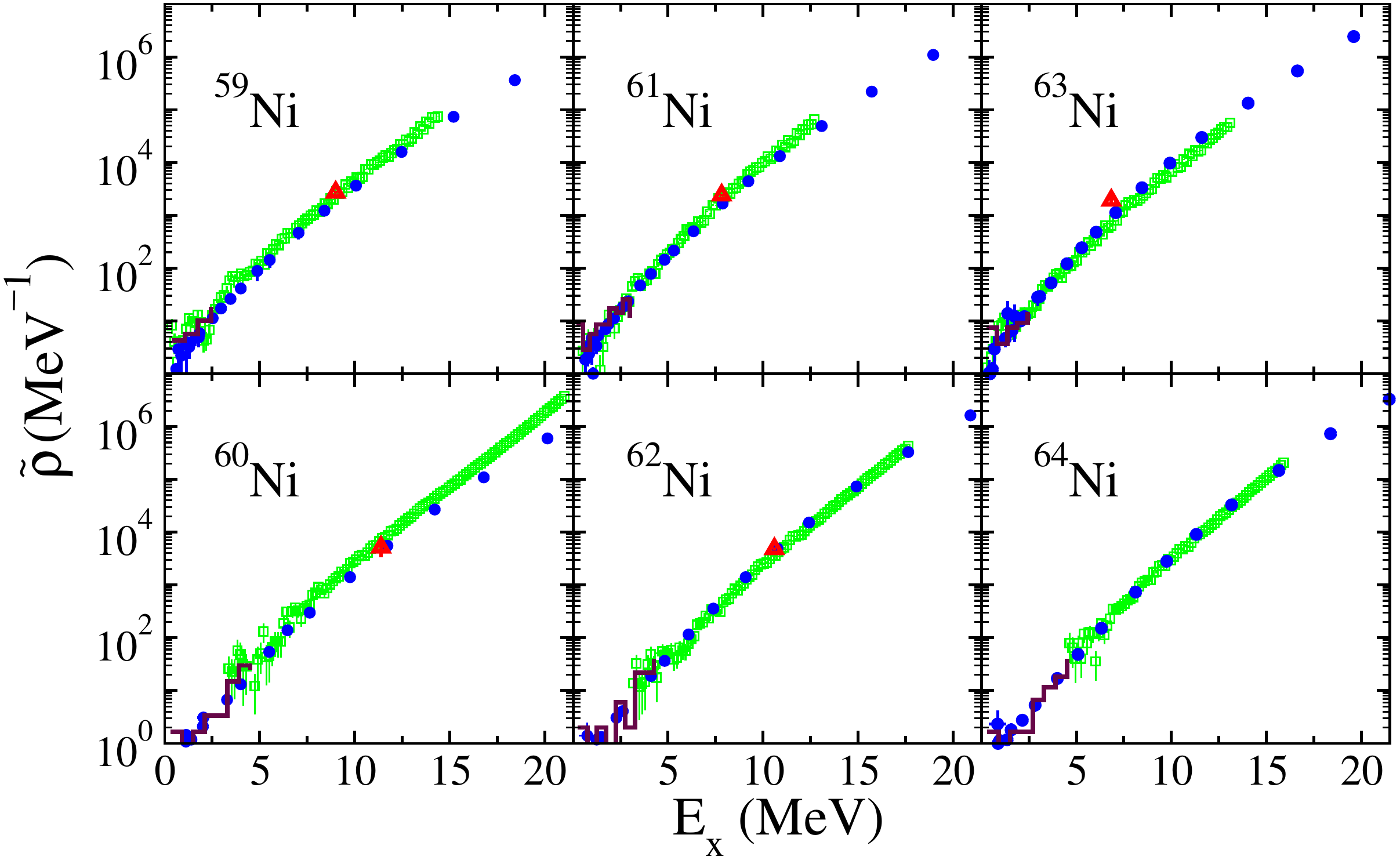}
}
\caption{Level densities $\tilde \rho$ of nickel isotopes $^{59-64}$Ni versus excitation energy $E_x$. The SMMC results (solid circles) are compared with level counting data at low excitation energies (solid histograms)~\cite{Ripl3}, neutron resonance data when available (triangles)~\cite{Ripl3} and level densities determined from proton evaporation spectra~\cite{Voinov2012,Voinov2012a} (open squares and quasi-continuous lines). Taken from Ref.~\cite{Bonett2013}.  }
\label{nickel}      
\end{figure*}

We discuss a recent application to the calculation of level densities in a family of nickel isotopes~\cite{Bonett2013} using the Hammiltonian of Ref.~\cite{Nakada1997} in the $fpg_{9/2}$ shell.  The level density $\tilde \rho$  is defined without counting the $2J+1$ magnetic degeneracy of levels with spin $J$, i.e., $\tilde \rho(E_x) = \sum_J \rho_J(E_x)$ where $\rho_J$ is the density of levels with spin $J$. Recent experiments extracted the level densities of the $^{59-64}$Ni  isotopes through the measurements of proton evaporation spectra~\cite{Voinov2012}. 

In SMMC, we can calculate the level density directly~\cite{Alhassid2015a} by projecting on the spin component $M$. Denoting by $\rho_M$ the density for levels with $J_z=M$, we have $\tilde\rho=\rho_{M=0}$ for even-mass nuclei and $\tilde\rho=\rho_{M=1/2}$ for odd-mass nuclei.

In Fig.~\ref{nickel}, we show the level densities for a family of nickel isotopes $^{59-64}$Ni. The ground-state energies of the odd-mass isotopes were determined accurately using our solution to the odd-particle sign problem discussed in Sec.~\ref{Odd-A}.

\section{Heavy nuclei}

Open shell heavy nuclei can be strongly deformed and are characterized by rotational collectivity.  An important question we discuss in Sec.~\ref{collectivity} is whether such deformed nuclei can be described in a spherical shell model approach using a truncated single-particle space.

 The SMMC method was successfully extended to heavy nuclei~\cite{Alhassid2008,Ozen2013}. The single-particle model space we use for calculations in rare-earth nuclei is composed of the $50-82$ shell plus the $1f_{7/2}$ orbital for protons, and the $82-126$ shell plus  the $0h_{11/2}, 1g_{9/2}$ orbitals for neutrons.  The corresponding many-particle model space for $^{162}$Dy is of order $\sim 10^{29}$.  

\begin{figure*}[bth]
\center\resizebox{0.7\textwidth}{!}{%
  \includegraphics{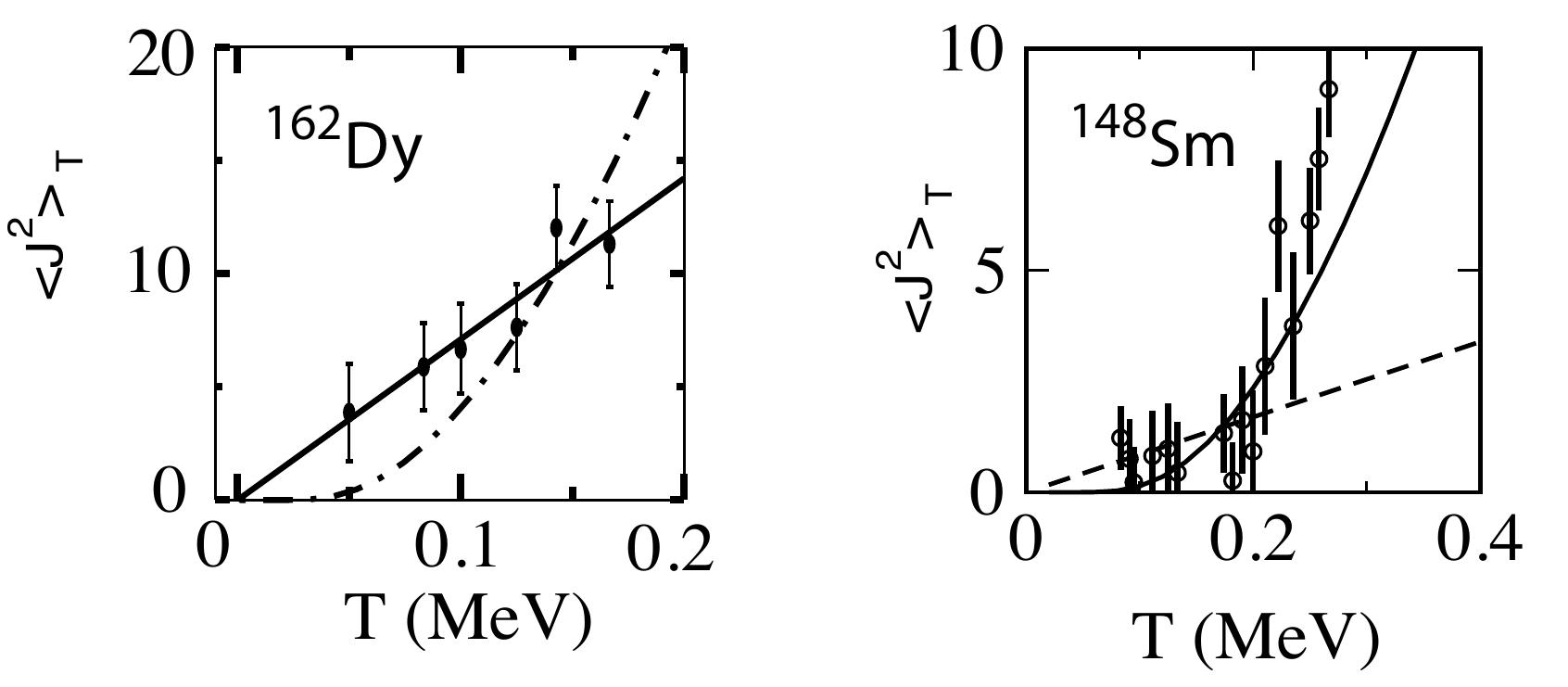}
}
\caption{Left: $\langle \mathbf{J}^2 \rangle_T$  in $^{162}$Dy. The SMMC results (solid circles) are compared with fits to the rotational model (solid line) and vibrational model (dashed-dotted line).  Adapted from Ref.~\cite{Alhassid2008}.  Right: $\langle \mathbf{J}^2 \rangle_T$  in $^{148}$Sm. The SMMC results (open circles) are compared with fits to the vibrational model (solid line) and rotational model (dashed line). Taken from Ref.~\cite{Alhassid2015}.}
\label{J2}      
\end{figure*} 

\subsection{Crossover from vibrational to rotational collectivity in the CI shell model approach}\label{collectivity}

 \begin{figure*}[bth]
\center\resizebox{0.85\textwidth}{!}{%
  \includegraphics{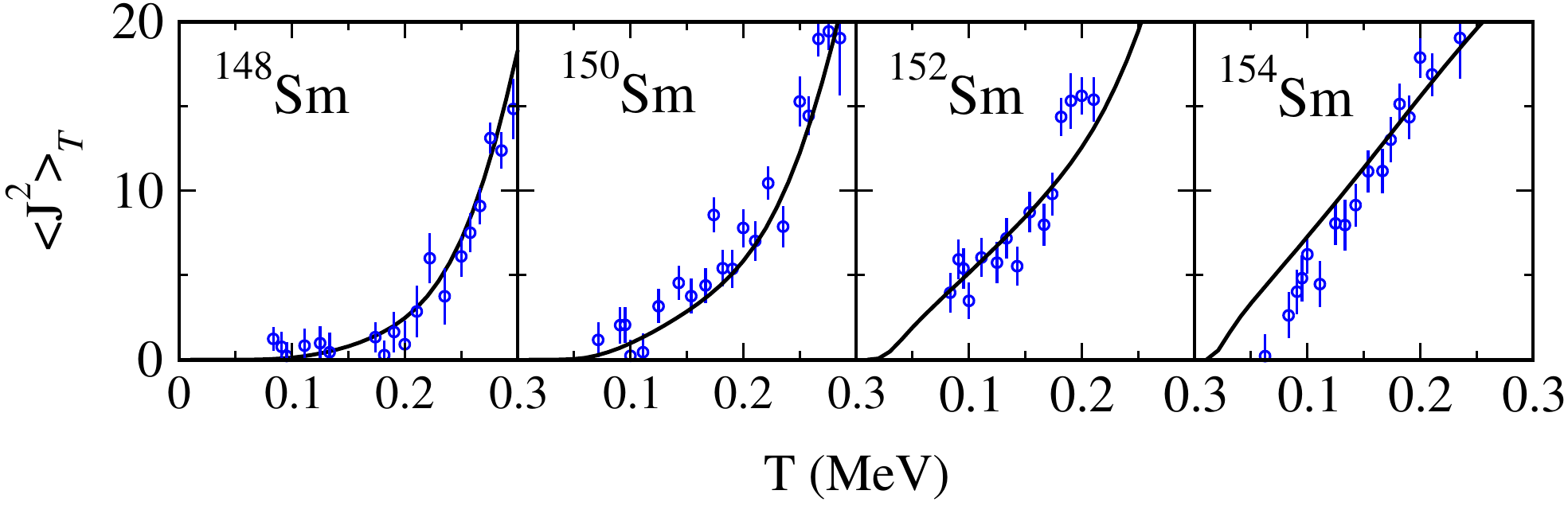}
}
\caption{$\langle \mathbf{J}^2 \rangle_T$ for a family of samarium isotopes $^{148-154}$Sm. The SMMC results (open circles with statistical errors) are in good agreement with values extracted from experimental data (solid lines; see text). Adapted from Ref.~\cite{Ozen2013}.}
\label{Sm-J2}      
\end{figure*}

\begin{figure*}[bth]
\center\resizebox{0.95\textwidth}{!}{%
  \includegraphics{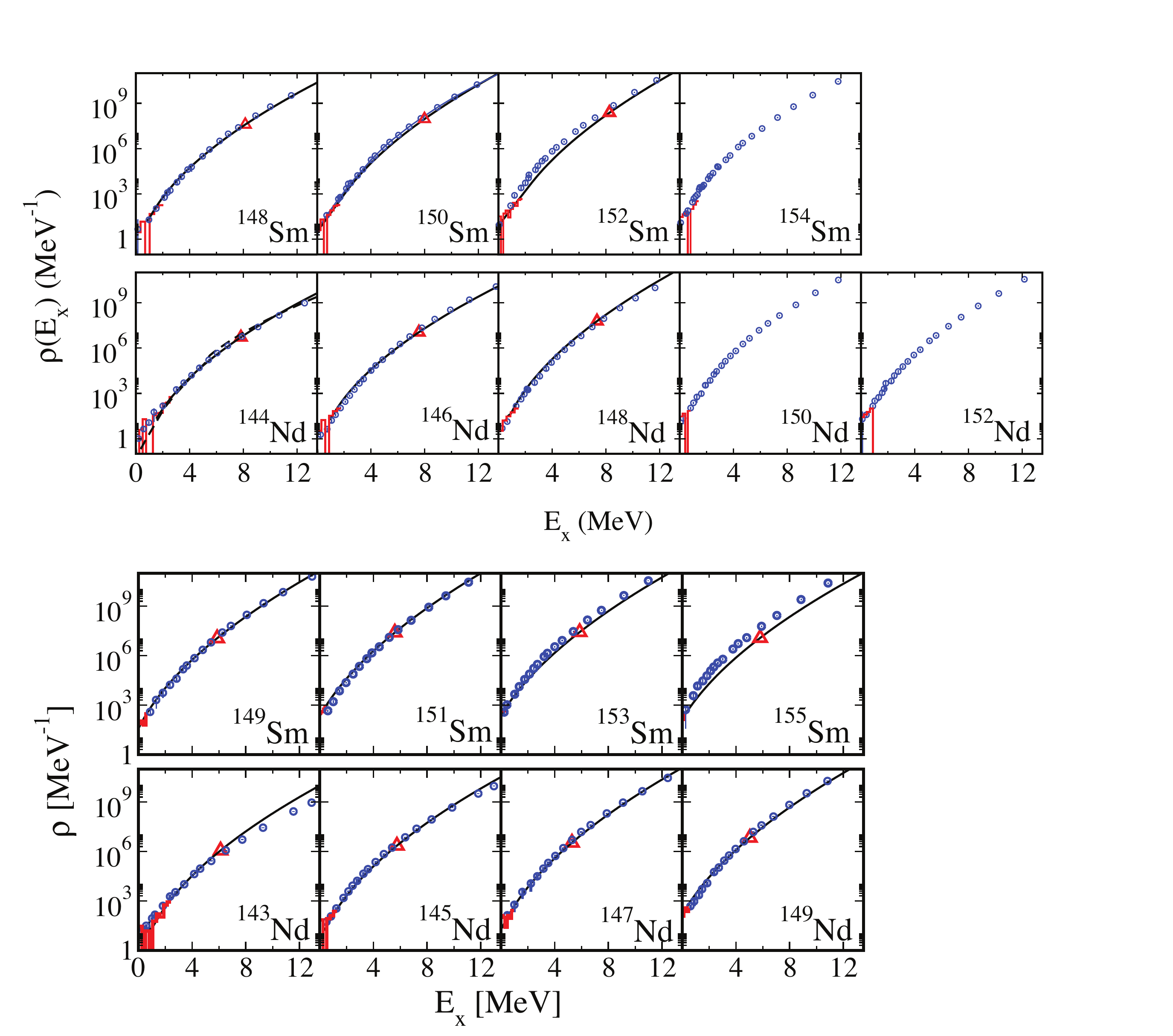}
}
\caption{State densities $\rho$ versus excitation energy $E_x$ in families of even-mass (top two panels) and odd-mass (bottom two panels) samarium and neodymium isotopes. The SMMC densities (open circles) are compared with level counting data at low excitation energies (solid histograms) and  neutron resonance data (triangles) when available~\cite{Ripl3}. The solid lines are  back-shifted Fermi gas densities extracted from experimental data. Adapted from Refs.~\cite{Ozen2013,Ozen2015,Alhassid2014a}.}
\label{rho}      
\end{figure*}

\begin{figure*}[bth]
\center\resizebox{0.9\textwidth}{!}{%
  \includegraphics{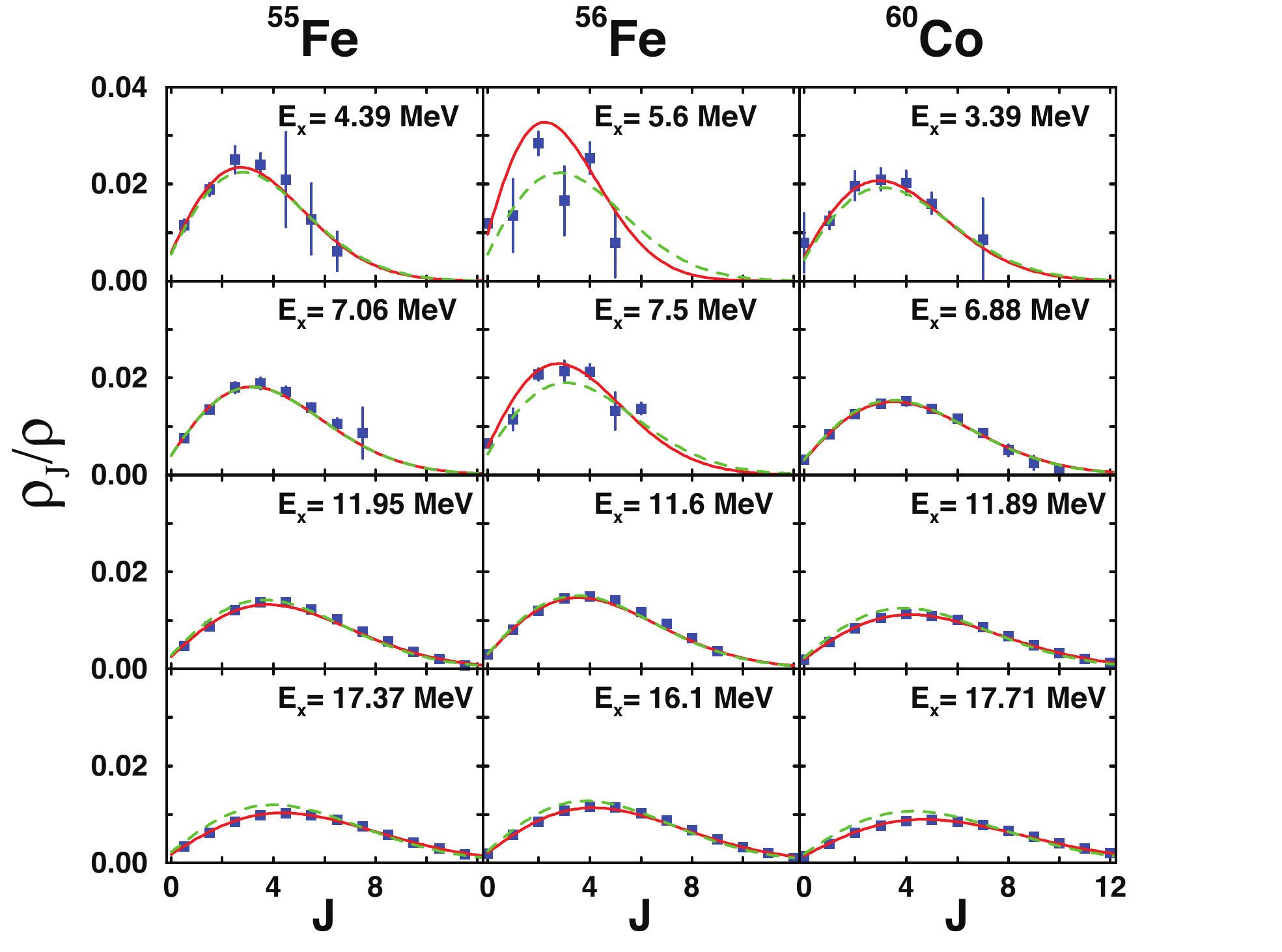}
}
\caption{Spin distributions $\rho_J/\rho$ versus spin $J$ for several values of the excitation energy $E_x$ and for the nuclei $^{55}$Fe (left column), $^{56}$Fe (middle column) and $^{60}$Co (right column). The SMMC results (solid squares with error bars) are compared with the spin cutoff model with a fitted spin cutoff parameter $\sigma$ (solid lines) and the spin cutoff model with rigid-body moment of inertia (dashed lines). Taken from Ref.~\cite{Alhassid2007}.}
\label{spin_dist}      
\end{figure*}

\begin{figure*}[bth]
\center\resizebox{0.8\textwidth}{!}{%
  \includegraphics{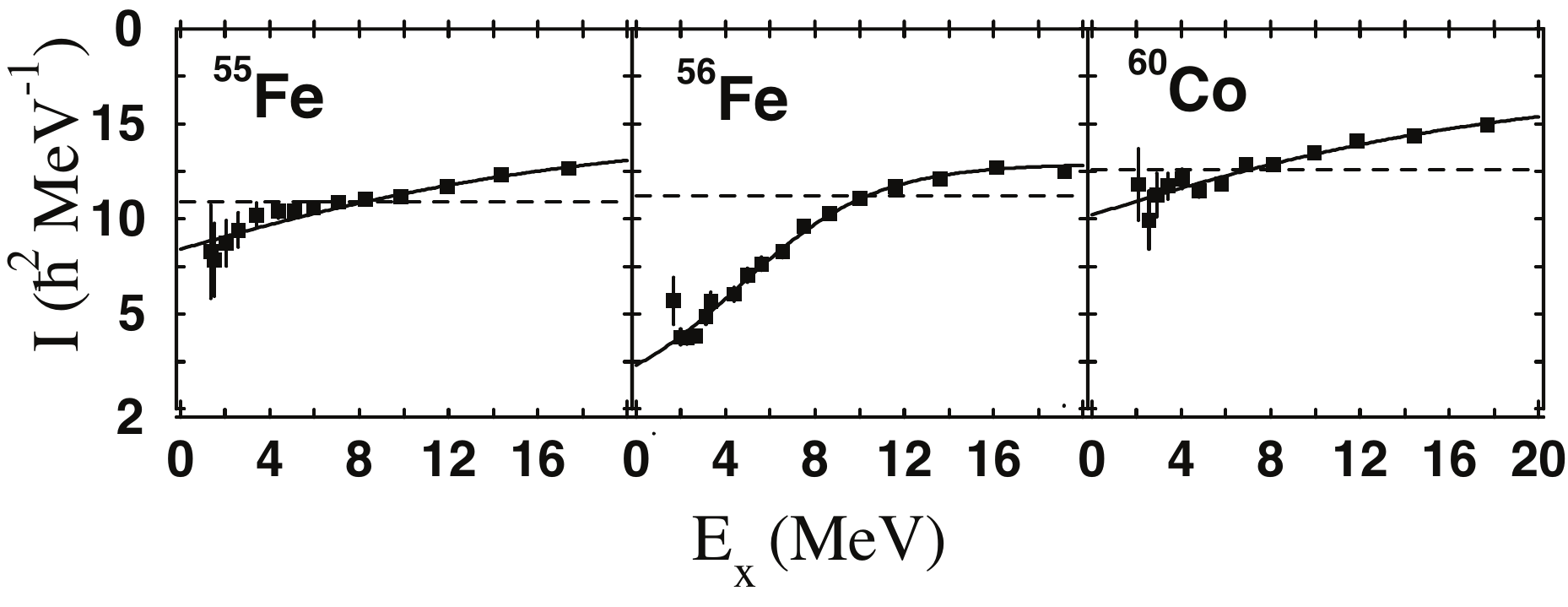}
}
\caption{Moment of inertia $I$ versus excitation energy $E_x$ for the nuclei $^{55}$Fe, $^{56}$Fe and $^{60}$Co. The moment of inertia $I$ extracted from $\sigma^2$ using Eq.~(\ref{sigma-I}) (solid squares with error bars) is compared with the rigid-body moment of inertia (dashed lines). Adapted from Ref.~\cite{Alhassid2007}.}
\label{inertia}      
\end{figure*}

The various types of collectivity, e.g., vibrational and rotational collectivity, are usually identified by their corresponding spectra. However,  SMMC does not provide detailed spectroscopy; it is more suitable for the calculation of thermal observables.  We have therefore identified an observable whose low-temperature behavior is sensitive to the type of collectivity. Such observable is $\langle {\bf J}^2\rangle_T$.  Its temperature dependence at low $T$ is given by~\cite{Alhassid2008,Ozen2013} 
\begin{eqnarray}\label{J2-theory}
\langle \mathbf{J}^2 \rangle_T \approx
 \left\{ \begin{array}{cc}
 30 { e^{-E_{2^+}/T} \over \left(1-e^{- E_{2^+}/T}\right)^2} &{\rm vibrational\; band}  \\
 \frac{6}{E_{2^+}} T & {\rm rotational \;band}
 \end{array} \right.
\end{eqnarray}
where $E_{2^+}$ is the excitation energy of the first $2^+$ level. In Fig.~\ref{J2} we show $\langle \mathbf{J}^2 \rangle_T$ versus $T$ for the deformed nucleus $^{162}$Dy (left panel) and for the spherical nucleus $^{148}$Sm (right panel). The solid circles with error bars are the SMMC results.  In $^{162}$Dy, they are well fitted by the rotational band model, while in $^{148}$Sm, the vibrational band model is a better fit. These results  validate the rotational character of $^{162}$Dy and the vibrational nature of $^{148}$Sm.

 The observable $\langle \mathbf{J}^2 \rangle_T$ also describes well the crossover from vibrational to rotational collectivity.  In Fig.~\ref{Sm-J2}, we show by solid circles the SMMC values of $\langle \mathbf{J}^2 \rangle_T$ as a function of temperature for a family of samarium isotopes $^{148-154}$Sm.  We observe a crossover from a ``soft'' response to temperature in $^{148}$Sm characterizing spherical vibrational nuclei, to a ``rigid'' linear response to temperature in  $^{154}$Sm characterizing deformed rotational nuclei. 
 
 The solid lines in Fig.~\ref{Sm-J2} are extracted from the experimental data using
 \begin{eqnarray}
\label{Eq:J2high}
 \langle \mathbf{J}^2 \rangle_T  =  \frac{1}{Z(T)} \left(\sum_i^N J_i(J_i+1)(2J_i+1)e^{-E_{i}/T}  \right. \nonumber \\
    \left. + \int_{E_{N}}^\infty d E_x \: \rho(E_x) \: \langle \mathbf{J}^2 \rangle_{E_x} \; e^{-E_x/T} \right)\;,
\end{eqnarray}
where $Z(T)=\sum_{i}^{N} (2J_i+1) e^{-E_i/T} + \int_{E_{N}}^\infty d E_x \rho(E_x) e^{-E_x/T}$ is the experimental partition function. The summations in Eq.~(\ref{Eq:J2high}) and in the expression for $Z(T)$ extend over a complete set of experimentally known low-lying levels $i$ with excitation energy $E_i$  and spin $J_i$, and $\rho(E_x)$ is the back-shifted Bethe formula (BBF) with parameters that are determined from level counting and neutron resonance data.

\subsection{State densities in families of samarium and neodymium isotopes}

The SMMC state densities of the even samarium isotopes $^{148-154}$Sm and the even neodymium isotopes $^{144-152}$Nd, calculated by the method described  in Sec.~\ref{density}, are shown in the top two rows of Fig.~\ref{rho} (open circles)~\cite{Ozen2013,Alhassid2014a}.  We find good agreement with experimental data including level counting at low excitation energies (solid histograms) and, when available, neutron resonance data (triangles) at the neutron binding energy~\cite{Ripl3}. The Hamiltonians used for the calculations are given in Ref.~\cite{Ozen2013}.

Recently, we calculated the state densities in odd-mass samarium ($^{149-155}$Sm) and neodymium ($^{143-149}$Nd) isotopes~\cite{Ozen2015}.  The corresponding densities are shown in the bottom two rows of Fig.~\ref{rho}. The odd-mass isotopes has an odd-particle sign problem. The solution discussed in Sec.~\ref{Odd-A} requires additional development for the heavy nuclei, in which much larger $\beta$ values are required to reach the ground-state energy $E_0$. We use experimental data to make a one-parameter fit  ($E_0$) of the SMMC thermal energy to the thermal energy calculated from the experimental data.  We find that the SMMC densities for the odd-mass isotopes are consistent with the experimental data using the same family of Hamiltonians as for the even-mass isotopes. 

\section{Projection methods}

To determine the density distributions at a given fixed value of an observable, we use projection methods by representing the corresponding Dirac $\delta$ function as a Fourier transform. Below we discuss projections on good quantum numbers, e.g., spin (Sec.~\ref{spin}), as well as on observables that do not commute with the Hamiltonian, e.g., the axial quadrupole operator $\hat Q_{20}$ (Sec.~\ref{quadrupole}).

\subsection{Spin distributions}\label{spin}

  We can project on $\hat J_z$, the spin component along the z axis, by using (within the HS representation) a Fourier transform decomposition for $\delta(\hat J_z - M)$~\cite{Alhassid2007}
\begin{equation}
  \label{M-project}
  {\rm Tr}_M \,U_\sigma = {1 \over 2J_s + 1} \sum\limits_{k =-J_s}^{J_s}
  e^{-i \varphi_k M} {\rm Tr}\,\left( e^{i\varphi_k \hat J_z} U_\sigma \right) \;.
\end{equation}
Here $J_s$ is the maximal many-particle spin in the model space and  $\varphi_k \equiv \pi {k / (J_s+1/2)}$  with $k=-J_s, \ldots J_s$ are quadrature points. 

 Using ${\rm Tr}_J \hat X= {\rm Tr}_{M=J} \hat X  - {\rm Tr}_{M=J+1} \hat X$ (which hold for any scalar operator $\hat X$), we can also calculate $\rho_J(E_x)$, i.e., the level density for a given spin  $J$. 

In Fig.~\ref{spin_dist} we show the spin distribution $\rho_J/\rho$ as a function of $J$ at different values of the excitation energies $E_x$ for three  mid-mass nuclei: the odd-even nucleus $^{55}$Fe, the even-even nucleus $^{56}$Fe and the odd-odd nucleus $^{60}$Co. We compare the SMMC results with the spin cutoff model~\cite{Ericson1960}
\begin{equation}
  \label{spin-cutoff}
  \rho_J(E_x) = \rho(E_x)
  {(2J+1) \over 2\sqrt{2 \pi} \sigma^3} e^{-{J(J+1) \over 2
      \sigma^2}}\;,
\end{equation}
where $\sigma$ is an energy-dependent spin-cutoff parameter. The spin-projected density $\rho_J(E_x)$  does not include the $2J+1$ magnetic degeneracy and are thus normalized according to  $\sum_J (2J+1) \rho_J(E_x) \approx \rho(E_x)$.  The distribution (\ref{spin-cutoff}) corresponds to a normal distribution in the angular momentum vector $\vec J$.   The solid lines in Fig.~\ref{spin_dist} describe the spin cutoff model distributions in which $\sigma$ is determined by a fit to the corresponding SMMC spin distributions. 

We observe that at higher excitation energies the spin cutoff model works quite well and this seems to be the case also at low energies for odd-even and odd-odd nuclei. However, for the even-even nucleus $^{56}$Fe, we observe at low excitation energies an odd-even staggering in spin. This phenomenon is generic for even-even nuclei and was observed in empirical analysis of low-lying energy levels across the table of nuclei~\cite{vonEgidy2008,vonEgidy2009}.

 The spin cutoff parameter is related to the thermal moment of inertia $I$ by 
\begin{equation}\label{sigma-I}
\sigma^2 = {I T \over \hbar^2 }  \;.
\end{equation}
 In Fig.~\ref{inertia} we show the moment of inertia $I$ versus excitation energy $E_x$ for the same three nuclei as in Fig.~\ref{spin_dist}. We find that at low excitation energies and for the even-even nucleus $^{56}$Fe, $I$ is suppressed to values that are below its rigid-body value because of pairing correlations.  In contrast, the moments of inertia for the odd-even nucleus $^{55}$Fe and the odd-odd nucleus $^{60}$Co are not reduced much below their respective rigid-body values. 

\subsection{Quadrupole distributions}\label{quadrupole}

The level density as a function of deformation is important input for modeling of shape dynamics such as nuclear fission. Here we review the first step toward such a calculation in the framework of the CI shell model approach~\cite{Alhassid2014}.

The axial quadrupole operator in the laboratory frame $\hat Q_{20}$ and the Hamiltonian $H$  do not commute,  and thus cannot be diagonalized simultaneously.  The distribution of  $\hat Q_{20}$ at temperature $T$ is then given by
\begin{equation}\label{prob1}
P_T(q) = \sum_n \delta(q - q_n) \sum_m \langle q,n|e,m\rangle^2 e^{-\beta
e_m} \;,
\end{equation} 
where $|q,n\rangle$ are eigenstates of $\hat Q_{20}$ satisfying ${\hat
 Q_{20}}|q,n\rangle = q_n |q,n\rangle$ and similarly 
$|e, m\rangle$ are eigenstates of $H$. 
We calculate $P_T(q)$ by using its Fourier representation 
\begin{equation}
\label{prob}
P_T(q)={1\over {\rm Tr}\,  e^{-\beta \hat H}} \int_{-\infty}^\infty
{d \varphi \over 2 \pi} e^{-i \varphi q }\, {\rm Tr}\, \left(e^{i \varphi \hat Q_{20}} e^{-\beta
\hat H} \right).
\end{equation} 
together with the HS transformation for $e^{-\beta H}$.  In practice, we approximate the quadrupole-projected trace by a discrete Fourier decomposition. We take an interval $[-q_{\rm max},q_{\rm max}]$ and divide it into $2L+1$ 
equal intervals of length $\Delta q=2q_{\rm
max}/(2L+1)$.  We have
\begin{equation}\label{fourier-q}
{\rm Tr}\left(\delta(\hat Q_{20} - q_m) U_\sigma \right) \!\!  \approx \!\! {1\over 2 q_{\rm max}}\! \! \sum_{k=-L}^L
  \!\!\!  e^{-i \varphi_k q_m} {\rm Tr}(e^{i \varphi_k \hat Q_{20}} \hat U_\sigma) \;,
\end{equation}
where $q_m=m \Delta q$ ($m=-L,\ldots,L$) and $\varphi_k = \pi k/q_{\rm max}$ ($k=-L,\ldots, L$).

\begin{figure}[]
\resizebox{0.45\textwidth}{!}{%
  \includegraphics{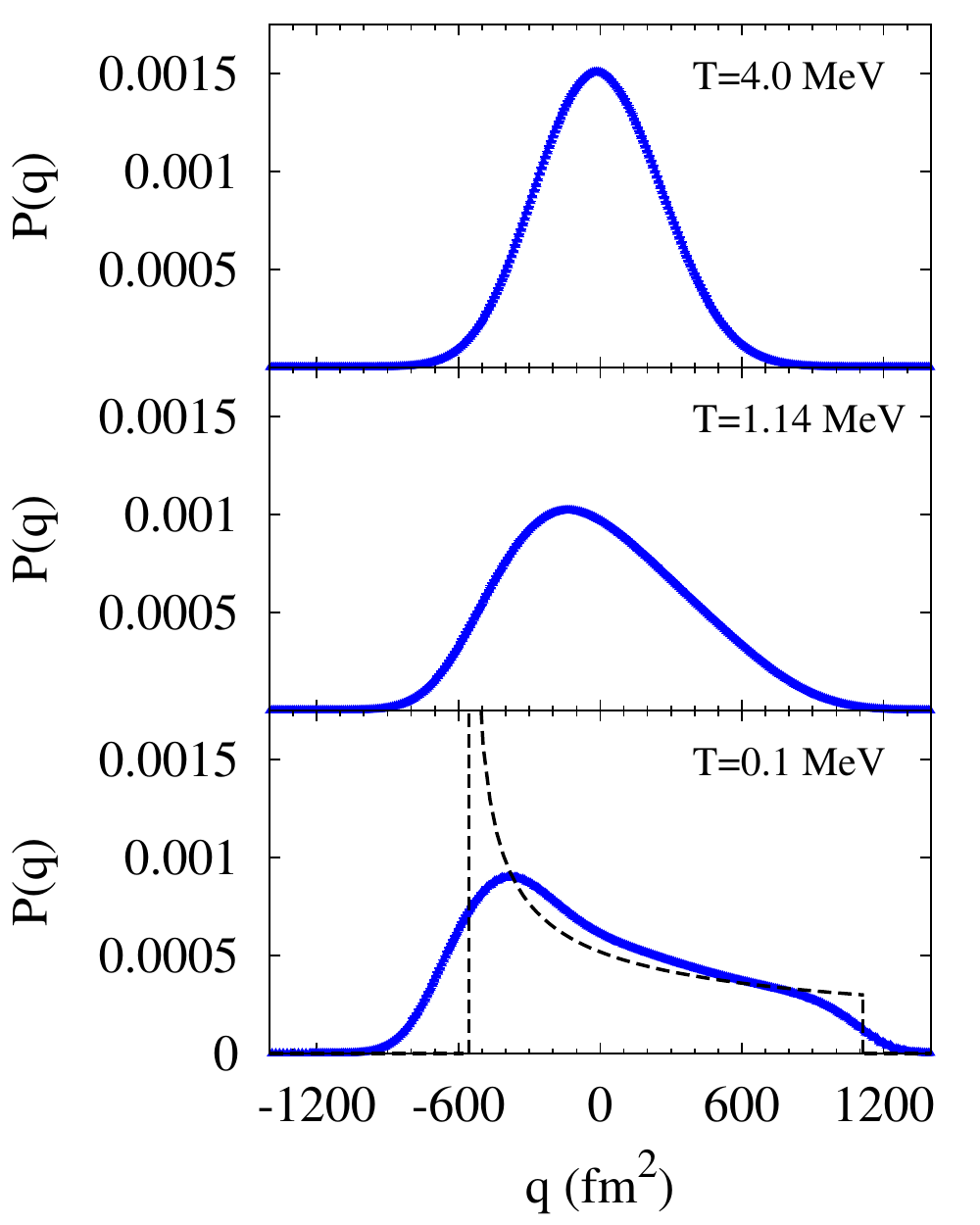}
}
\caption{Axial quadrupole distributions $P_T(q)$ in the laboratory frame for the deformed nucleus $^{154}$Sm  at three temperatures. The dashed line at the low temperature $T=0.1$ MeV is the quadrupole distribution of a prolate rigid rotor. Taken from Ref.~\cite{Alhassid2014}.}
\label{q-dist-sm154}      
\end{figure}

The SMMC quadrupole distributions $P_T(q)$ are shown in Fig.~\ref{q-dist-sm154} for the deformed nucleus $^{154}$Sm at several temperatures. At the lowest temperature $T=0.1$ MeV, we observe a skewed distribution which we compare with the axial quadrupole distribution of a prolate rigid rotor (dashed line). The overall similarity of the distributions is a clear signature of deformation. Deformation is an important concept in our understanding of heavy nuclei but it is introduced in the framework of a mean-field approximation (e.g, Hatree-Fock) that breaks rotational symmetry. Here we observe a signature of deformation in the quadruple distribution in the laboratory frame using a rotationally invariant CI shell model approach.  

In the finite-temperature HF approximation, $^{154}$Sm undergoes a shape transition from a deformed to a spherical nucleus. The quadrupole distribution is still skewed in the vicinity of the transition temperature (middle panel in Fig.~\ref{q-dist-sm154}). At a high temperature of $T=4$ MeV, the distribution is close to a Gaussian (top panel). In $^{148}$Sm, a spherical nucleus, we find that the quadrupole distributions are close to a Gaussian even at low temperatures.

Our goal is to eventually determine the distribution of the intrinsic quadrupole deformation. To this end, we construct combinations of the second rank quadrupole operator $\hat Q_{2\mu}$ that are invariant under rotations.  For example, $\hat Q\cdot \hat Q$ is a second order invariant, and $(\hat Q\times \hat Q)^{(2)}\cdot \hat Q$ is a cubic invariant.  The expectation values of these invariants are independent of the particular frame they are evaluated in, and in particular, they can be evaluated in the intrinsic frame.  Effective values of the intrinsic deformation parameters $\beta$ and $\gamma$ can then be calculated from these invariants
\begin{eqnarray}
\beta = \frac{\sqrt{5 \pi}}{ 3 r_0^2 A^{5/3} } \langle \hat Q \cdot \hat Q \rangle^{1/2}  \nonumber \\
\cos 3\gamma = -\sqrt{7 \over 2} {\langle (\hat Q \times \hat Q ) \cdot \hat Q \rangle \over  \langle \hat Q \cdot \hat Q \rangle^{3/2} } \;.
\end {eqnarray}

To determine the intrinsic quadrupole shape distribution $P_T(\beta,\gamma)$, we use the fact that $P_T(\beta,\gamma)$ is a rotational invariant, and expand $-\ln P_T(\beta,\gamma)$ in the quadrupole invariants. Carrying the expansion up to the fourth order invariant, we have
\be\label{landau}
-\ln P_T(\beta,\gamma) = N + A\beta^2 -B \beta^3\cos 3\gamma + C \beta^4 + \ldots \;,
\ee
where $A,B, C$ are temperature-dependent parameters and $N$ is a normalization constant.  Eq.~(\ref{landau}) carries a similarity to the expansion of the free energy in the quadrupole shape order parameter in the Landau theory of shape transitions~\cite{Levit1984,Alhassid1986}. The parameters $A,B,C$ can be determined from the expectation values of the three quadrupole invariants $\beta^2,  \beta^3\cos 3\gamma$ and $\beta^4$.
   In Ref.~\cite{Alhassid2014} we showed that the expectation values of these  low-order quadrupole invariants can be related to moments of $\hat Q_{20}$ in the laboratory frame, and can therefore be extracted from the distributions $P_T(q)$.  Once the distributions $P_T(\beta,\gamma)$ are known, we can use the saddle-point approximation to convert them to the intrinsic shape distributions $P_{E_x}(\beta,\gamma)$ at fixed excitation energy $E_x$.

After we determine $P_{E_x} (\beta,\gamma) $, the joint level density distribution $\rho(\beta,\gamma,E_x)$ at a given intrinsic deformation $\beta,\gamma$ and excitation energy $E_x$  can be calculated  from
\begin{equation}
\rho(\beta,\gamma,E_x) = \rho(E_x) P_{E_x} (\beta,\gamma) \;.
\end{equation}
 
\section{Conclusion and prospects}

The SMMC method is a powerful method for the calculation of level densities in the presence of correlations in very large model spaces.  A recent method to circumvent the odd-particle sign problem enabled us to calculate accurately level densities in odd-mass nuclei.  We demonstrated that a spherical CI shell model approach defined within a truncated single-particle space is capable of describing microscopically various types of collectivity in heavy nuclei.  Projection methods enable the calculation of the distribution of good quantum numbers such as spin.  We can also project on observables that do not commute with the Hamiltonian, such as the axial quadrupole operator.  Using such a quadrupole projection, we showed that nuclear deformation can be studied in the rotationally invariant framework of the CI shell model. 

The dependence of level densities on intrinsic nuclear deformation is important input in the modeling of shape dynamics and, in particular, nuclear fission. Our recent study of deformation in the framework of the CI shell model can be viewed as the first step toward the calculation of level densities versus deformation.

In comparing the calculated SMMC densities with neutron resonance data, we converted the measured  $s$-wave neutron resonance mean spacing to total densities assuming a spin cutoff model with rigid-body moment of inertia and equal positive- and negative-parity densities. It would be interesting to make a direct comparison of the measured average neutron resonance density with the corresponding sum of SMMC spin-parity level densities that are allowed by the selection rules. This requires the use of a combined spin-parity projection method in SMMC.

In our SMMC studies of level densities we use CI shell model Hamiltonians that are specific to the mass region under consideration.   A long-term goal is to derive an effective shell model Hamiltonian from a density functional theory.  Since density functional theories are valid globally across the table of nuclei, this would enable us to derive systematically CI shell model Hamiltonians across different mass regions.  Preliminary ideas for how to construct a map from a density functional theory on a CI shell model Hamiltonian were discussed in Refs.~\cite{Alhassid2006,Guzman2008}.  
It would also be useful to extend the SMMC studies to heavier mass regions (e.g., actinides), and to unstable nuclei. 

\subsection{Acknowledgments}

I would like to thank G.F. Bertsch, M. Bonett-Matiz, L. Fang,  C.N. Gilbreth, S. Liu,  A. Mukherjee, H. Nakada, and C. \"Ozen for their collaboration on the work reviewed above. This work was supported in part by the DOE grant DE-FG-0291-ER-40608. The research presented here used resources of the National Energy Research Scientific Computing Center, which is supported by the Office of Science of the U.S. Department of Energy under Contract No.~DE-AC02-05CH11231.  It also used resources provided by the facilities of the Yale University Faculty of Arts and Sciences High Performance Computing Center.

%

\end{document}